\documentclass{ws-procs9x6}            
\usepackage{subcaption}
\usepackage{lineno}
\begin{document}
\title{Measurement of Longitudinal Spin Asymmetries for Weak Boson Production at STAR}

\author{Jinlong Zhang, for the STAR Collaboration}

\address{Department of Physics and Astronomy, Stonry Brook University,\\
Stony Brook, New York 11794, USA\\
E-mail: jinlong.zhang@stonybrook.edu}

\begin{abstract}
The production of $W^\pm$ bosons in longitudinally polarized proton-proton collisions at RHIC provides a direct probe for the spin-flavor structure of the proton through the parity-violating single-spin asymmetry, $A_L$. At STAR, the leptonic decay channel $W \to e\nu$ can be effectively measured with the electromagnetic calorimeters and time projection chamber. STAR has measured the $A_L(W)$ as a function of the decay-electron's pseudorapidity from datasets taken in 2011 and 2012, which has provided significant constraints on the helicity-dependent PDFs of $\bar{u}$ and $\bar{d}$ quarks. 
In 2013 the STAR experiment collected an integrated luminosity of $\sim$250 pb$^{-1}$ at $\sqrt{s}=510$ GeV with an average beam polarization of $\sim$56\%, which is  more than three times larger than the total integrated luminosity of previous years. The final results from 2013 dataset for $W$-boson $A_L$ will be reported. Also the impacts of STAR data on our knowledge of the sea-quark spin-flavor structure of the proton will be discussed.
\end{abstract}


\bodymatter

\section{Introduction}\label{aba:sec1}
The proton's spin structure has attracted both theoretical and experimental interest in the past few decades. Polarized inclusive deep-inelastic scattering (DIS) experiments have provided data showing that the quark and antiquark spins only contribute $\sim$30\% of the proton spin~\cite{sumR}. In semi-inclusive DIS measurements, the flavor decomposition of quark spin contribution to proton spin can be accessed by identifying one or more hadrons in the final state. Fragmentation functions are required to relate the final state hadrons to the scattered quarks and antiquarks. Uncertainties of the flavor separated quark and antiquark spin contributions are still relatively large~\cite{dssv}\cite{lss}. 

In unpolarized Drell-Yan experiments, a flavor asymmetry between $\bar{u}$ and $\bar{d}$ has been observed~\cite{Baldit1994, Towell2001}. It is natural to ask if such a flavor asymmetry also exists in the polarized sea. Different models developed to explain the unpolarized flavor asymmetry, however, gave very different predictions~\cite{Chang2014}. Pioneering measurements were made by COMPASS~\cite{compass} but limited by precision. So, experimental input from the RHIC spin program becomes critical.   

As one of the featured measurements of the RHIC spin program, $W^\pm$ boson production in polarized proton-proton collisions at RHIC were proposed as a unique tool to study the spin-flavor structure of the proton at a high scale, $Q\sim M_W$~\cite{rhic_spin, ma2017}, where $Q$ describes the exchanged energy. Due to the parity violation of the weak interaction, $W^{\pm}$ bosons only couple to left-handed quarks and right-handed antiquarks. They naturally determine the helicity of the incident quarks. The charge of $W$ boson selects a specific combination of the flavor of the incoming quarks, $u+\bar{d} \rightarrow W^+$ and $d+\bar{u} \rightarrow W^-$. Subsequently, their leptonic decays provide a fragmentation-function-free probe of the helicity-dependent Parton Distribution Functions (PDFs). 

The longitudinal single-spin asymmetry is defined as  $A_L=(\sigma_+-\sigma_-)/(\sigma_++\sigma_-)$, where $\sigma_{+(-)}$ is the cross section when the polarized beam has positive (negative) helicity. At leading order, the $A_L$ of $W^\pm$ are directly sensitive to $\Delta \bar{d}$ and $\Delta \bar{u}$,

\begin{equation} 
A_L^{W^+} \propto \frac{ \Delta \bar{d}(x_1) u(x_2) - \Delta u(x_1) \bar{d}(x_2) }{\bar{d}(x_1) u(x_2) + u(x_1) \bar{d}(x_2)},
\label{Eqn:ALWp}
\end{equation}
\begin{equation}
A_L^{W^-} \propto \frac{\Delta \bar{u}(x_1) d(x_2)-\Delta d(x_1) \bar{u}(x_2)}{\bar{u}(x_1)d(x_2) + d(x_1) \bar{u}(x_2)},
\label{Eqn:ALWn}
\end{equation} 
where $x_1$ and $x_2$ are the momentum fractions carried by the scattering partons. The $A_L^{W^+}$ ($A_L^{W^-}$) approaches $\Delta u/u$ ($\Delta d/d$) in the very forward region of $W$ rapidity, $y_W \gg 0$, and $-\Delta \bar{d}/\bar{d}$ ($-\Delta \bar{u}/\bar{u}$) in the very backward region of $W$ rapidity $y_W \ll 0$.  

First measurements of the $W$ single-spin asymmetry at RHIC were reported by STAR\cite{PRLW} and PHENIX\cite{PhenixW} collaborations from data collected during a successful commission run at $\sqrt{s}$ = 500 GeV in 2009. In the following proton-proton running years, both STAR\cite{PRL-2014} and PHENIX\cite{PHenixW2016} performed further measurements of $W$ $A_L$ with increased statistics and improved beam polarization. In 2013, STAR collected an integrated luminosity of $\sim$250 pb$^{-1}$ at $\sqrt{s}$ = 510 GeV with an average beam polarization of $\sim$56\%. This is  more than three times larger than the total integrated luminosity of previous years. In this contribution, we report the final results on $W$ $A_L$ from STAR data obtained in 2013 and the impact on flavor-separated light quark and antiquark polarization\cite{STAR2013}.

\section{Analysis}\label{aba:sec2}

STAR measures the decay electrons (positrons) in $W\rightarrow e\nu$. The Time Projection Chamber (TPC) covering the full azimuth and a pseudorapidity range of $-1.3 < \eta < 1.3$, is the main tracking subsystem. It provides momenta and charge sign information for charged particles. Outside the TPC, the Barrel and Endcap Electromagnetic Calorimeters (BEMC and EEMC) covering full azimuth and pseudorapidity ranges of $-1 < \eta < 1$ and $1.1 < \eta < 2.0$ respectively, measure the energy of electrons and photons. A $W^{\pm} \rightarrow e^{\pm}\nu$ candidate event is characterized by a well isolated electron track carrying transverse energy, $E_T^e$, which exhibits the two-body decay ``Jacobian Peak" near half of the $W^{\pm}$ mass, $\sim$40 GeV. The undetected decay neutrinos lead to a large missing energy in the opposite azimuth of the $e^\pm$ candidates, so there will be a significant $p_T$ imbalance when summing over all reconstructed final-state objects. In contrast, the $p_T$ vector is well balanced for  background events such as $Z/\gamma^* \rightarrow e^+e^-$ and QCD di-jet or multi-jet events. The $W$ selection is achieved based on these isolation and $p_T$ imbalance features. 

STAR is not a 4$\pi$ coverage detector. A di-jet event or $Z/\gamma^* \rightarrow e^+e^-$ event could have one of its jets or electrons outside the STAR acceptance. 
Such an event could be accepted if the detected jet or electron passes all the $W$ selection criteria. In addition, a $W$ boson can decay to $\tau+\nu$ and $\tau$ can further decay to electrons. We can not distinguish these feed down electrons from signal electrons. Contributions from $Z/\gamma^*$ and $\tau$ which are well understood, are estimated from Monte Carlo (MC) simulation including all detector and luminosity effects. 
The QCD background is estimated using two procedures. The existing EEMC is used to assess the background from the corresponding uninstrumented acceptance region on the opposite side of the collision point. The remainder of the QCD background is estimated by normalizing the $E_T$ spectrum of an pure QCD sample to the observed $E_T$ spectrum in a QCD dominated interval. 

\section{Results}\label{aba:sec3}
From the spin sorted yields of $W^{\pm}$ bosons, the longitudinal single-spin asymmetries were extracted in four pseudorapidity intervals, using
\begin{equation} 
A_L = \frac{1}{\beta}\frac{1}{P}\frac{N_+/l_+ - N_-/l_-}{N_+/l_+ + N_-/l_-}, 
\label{eq:AL}
\end{equation}
where $\beta$ quantifies the dilution due to background, $P$ is the beam polarization, $N_+(N_-)$ is the $W$ yield when the helicity of the polarized beam is positive (negative), and $l_{\pm}$ are the relative luminosity correction factors. 

\begin{figure}[t]
\centering
\begin{subfigure}{0.49\textwidth}
  \centering
  \includegraphics[width=\linewidth]{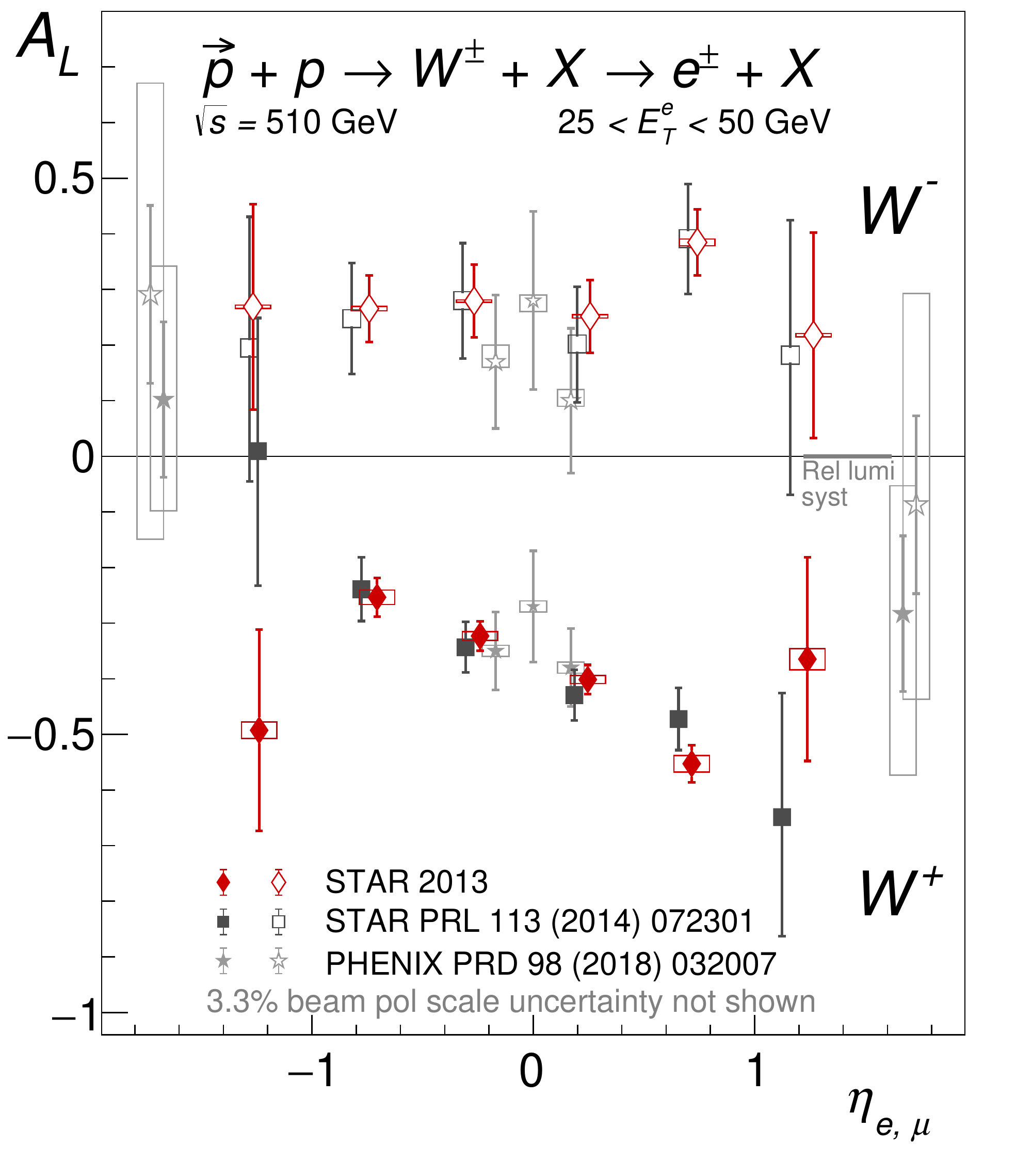}  
  \caption{Final 2013 results}
  \label{Fig:WAL1}
\end{subfigure}
\begin{subfigure}{0.49\textwidth}
  \centering
  \includegraphics[width=\linewidth]{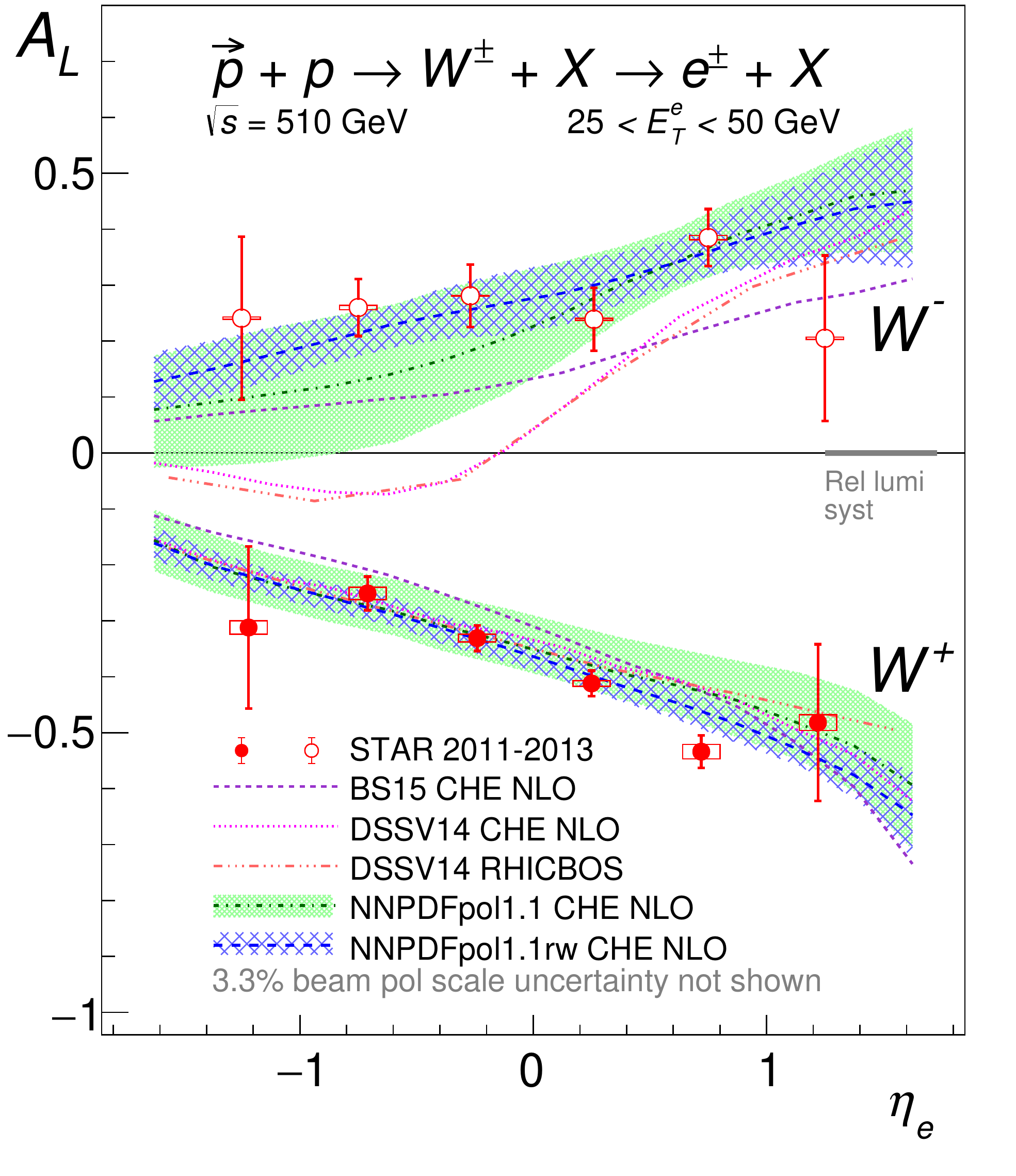}  
  \caption{Combined 2011-2013 results}
  \label{Fig:WAL2}
\end{subfigure}
\caption{Longitudinal single-spin asymmetry, $A_L$, for $W^\pm$ production as a function of the lepton pseudorapidity, $\eta_{e}$, in comparison to theory predictions. \label{Fig:WAL} }
\end{figure}

From STAR 2011+2012 $W^\pm$ $A_L$ results, it was noted that a larger $\Delta\bar{u}$ is preferred. 
With $\sim$40\% smaller uncertainties, the 2013 results confirmed the larger $\Delta\bar{u}$ preference, as shown in Fig~\ref{Fig:WAL1} by the red data points. The combined 2011+2012 and 2013 results are shown in Fig~\ref{Fig:WAL2} in comparison to the theory predictions. 
The 2011+2012 results have been included into the global QCD analysis by NNPDF group~\cite{nnpdf}. The constraints provided by these STAR data lead to a shift in the central value of $\Delta\bar{u}$ from negative to positive for 0.05 $< x <$ 0.25, which RHIC is sensitive to. The data favor $\Delta \bar{u} > \Delta\bar{d}$ which is opposite to the unpolarized distributions.  

\begin{figure}[ht]
\centering
\begin{subfigure}{0.32\textwidth}
  \centering
  \includegraphics[width=\linewidth]{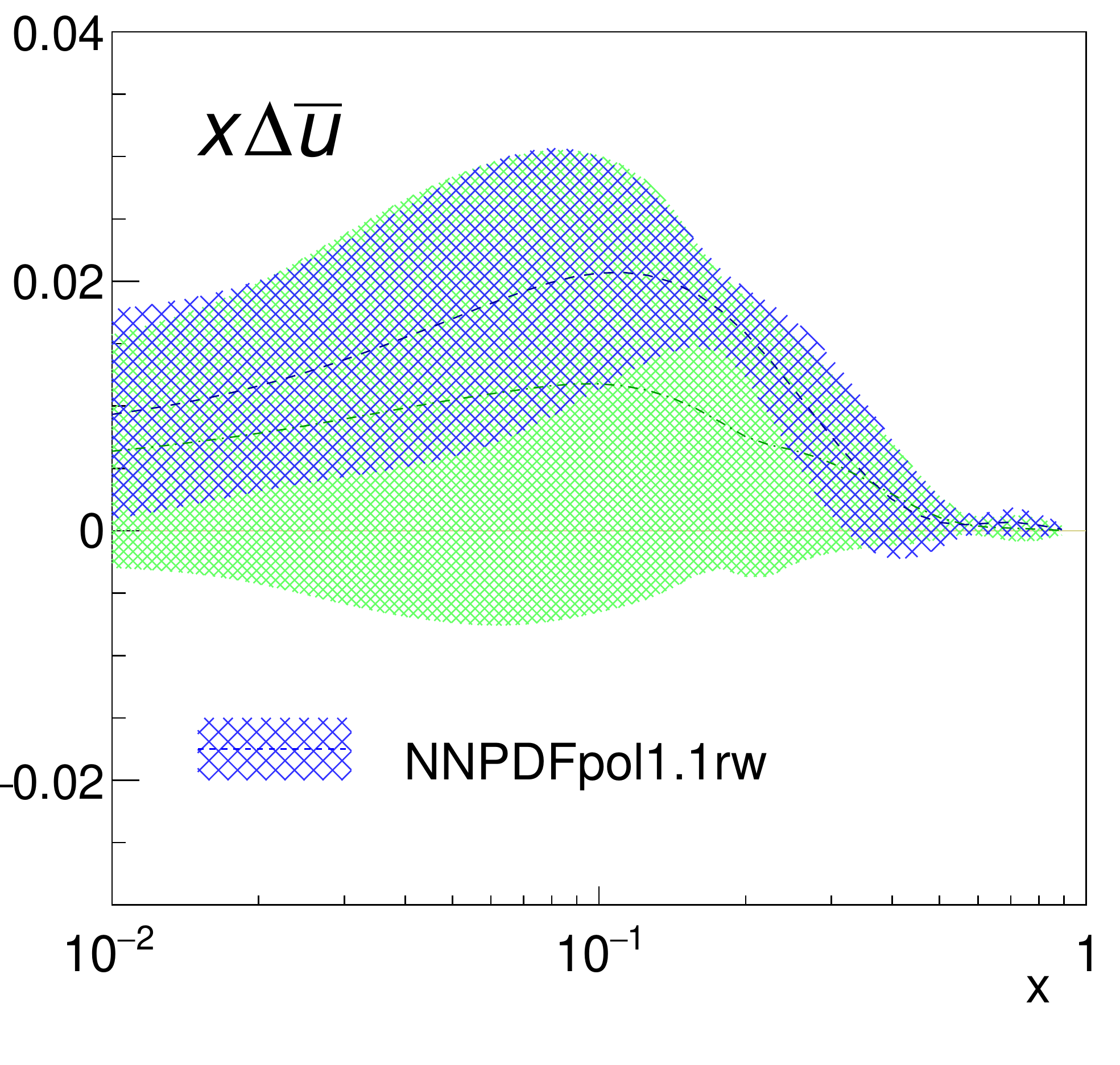}  
  \caption{$\Delta\bar{u}$}
  \label{Fig:sea1}
\end{subfigure}
\begin{subfigure}{0.32\textwidth}
  \centering
  \includegraphics[width=\linewidth]{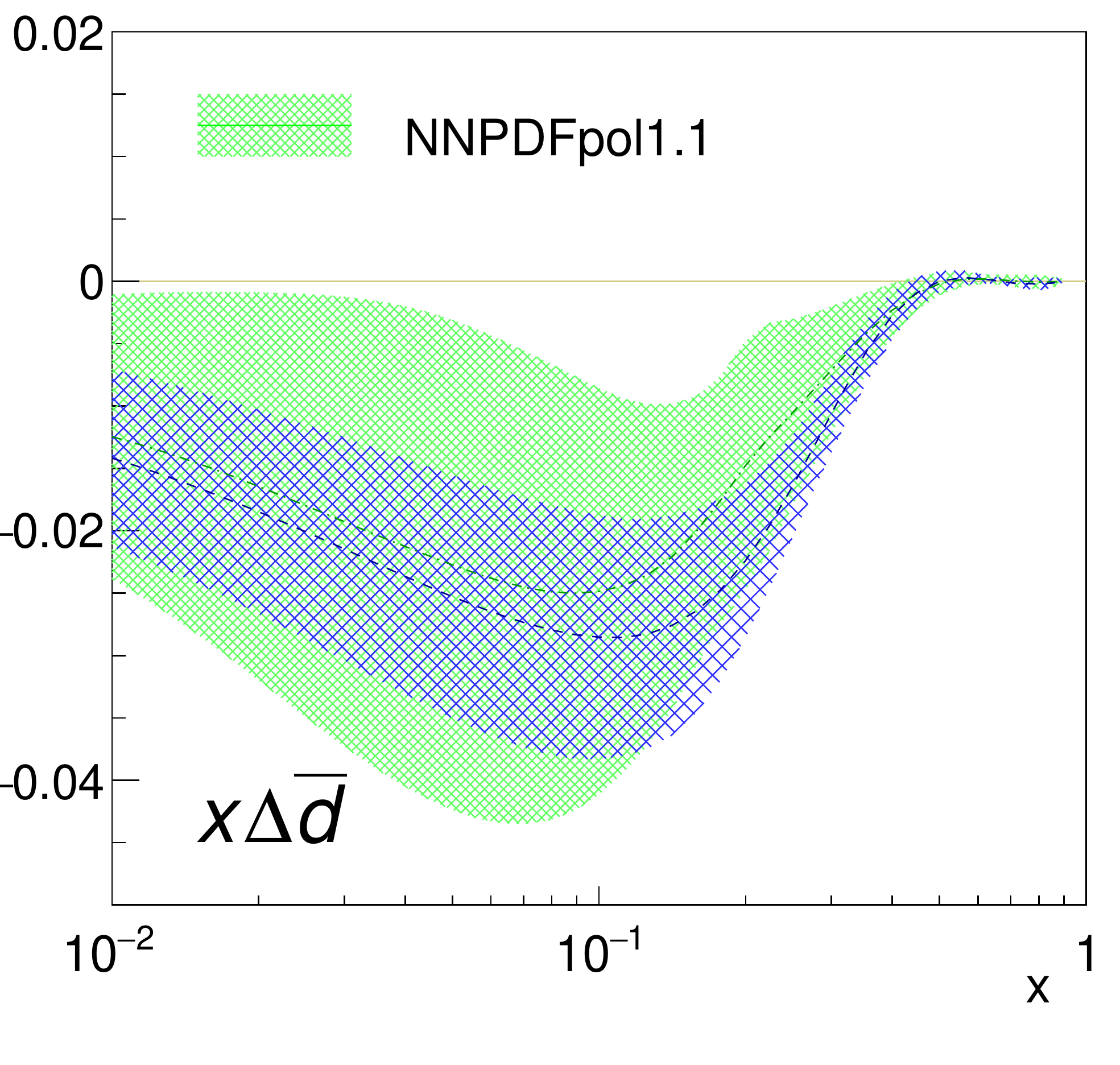}  
  \caption{$\Delta\bar{d}$}
  \label{Fig:sea2}
\end{subfigure}
\begin{subfigure}{0.32\textwidth}
  \centering
  \includegraphics[width=\linewidth]{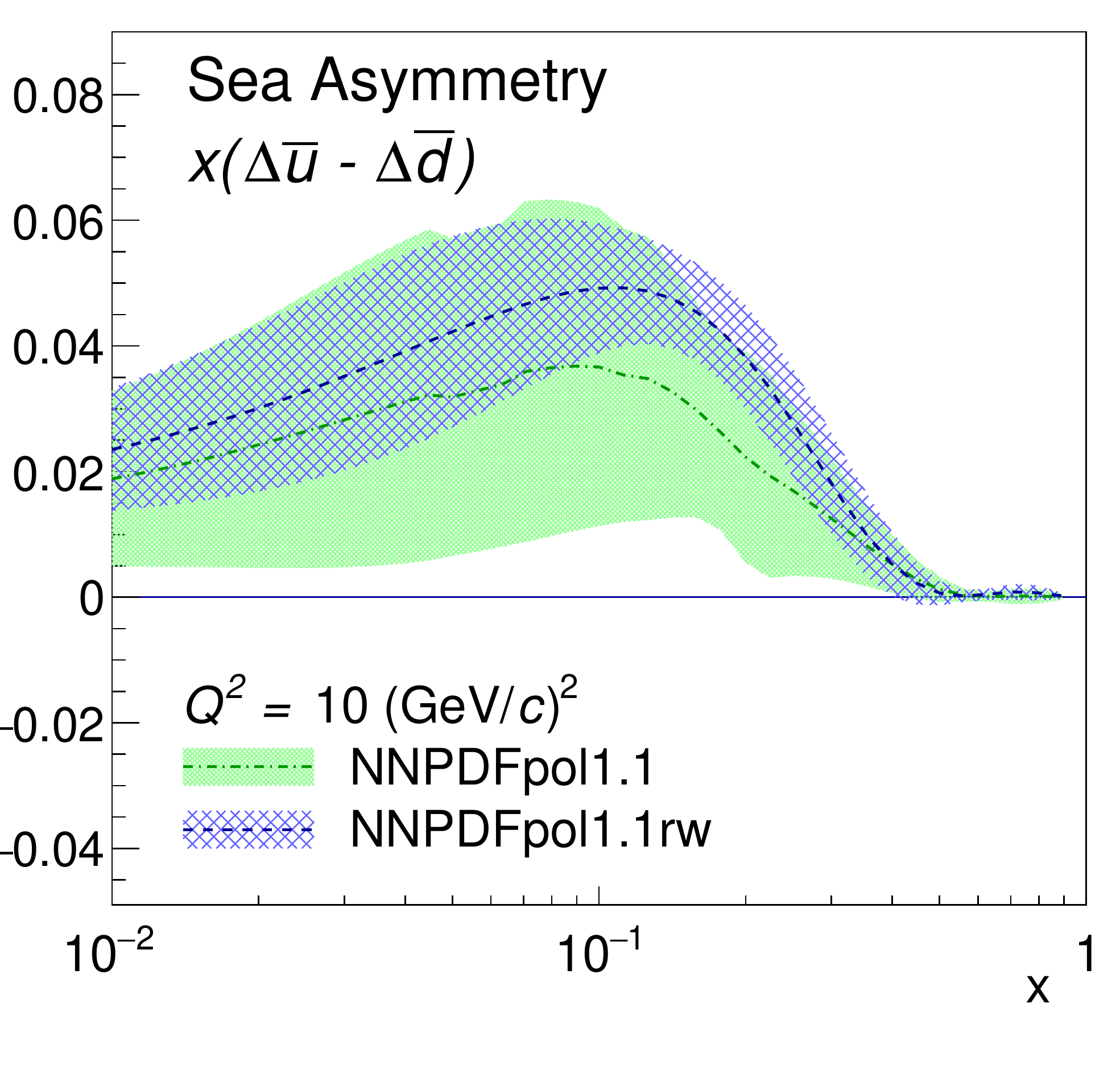}  
  \caption{$\Delta\bar{u} - \Delta\bar{d}$}
  \label{Fig:sea3}
\end{subfigure}
\caption{Impacts of STAR 2013 $W^\pm$ $A_L$ results on light sea polarizations and the flavor asymmetry between $\Delta\bar{u}$ and $\Delta\bar{d}$.\label{Fig:polsea} }
\end{figure}

To quantitatively assess the impacts of STAR 2013 $W^\pm$ $A_L$ results, a reweighting method~\cite{reweighting} was implemented with the 100 publicly available replicas of NNPDFpol1.1 PDFs. The reweighted $W^\pm$ $A_L$ predictions are shown in Fig~\ref{Fig:WAL2} as the blue hatched band. Correspondingly, the impacts on the $\bar{u}$ and $\bar{d}$ polarization are shown in Fig~\ref{Fig:sea1} and \ref{Fig:sea2}. The green and blue hatched bands are the distributions before and after reweighting with 2013 results respectively. Now, it can be found that $\Delta\bar{u}$ is positive and $\Delta\bar{d}$ is negative at medium $x$. And, the asymmetry between them, $\Delta\bar{u} - \Delta\bar{d}$ has similar size, but opposite sign compared to the flavor asymmetry of the unpolarized sea. 
 
\bibliographystyle{ws-procs9x6} 


\end{document}